\documentclass[5p,times]{elsarticle}
\usepackage{graphicx}
\usepackage{subcaption}
\usepackage{hyperref}
\usepackage{hyphenat}
\usepackage{microtype}
\usepackage{mathtools}
\usepackage[utf8]{inputenc}
\usepackage[T1]{fontenc}
\usepackage{xcolor}
\usepackage{arydshln} 

\usepackage{cleveref}

\biboptions{sort&compress}
\journal{Science Bulletin}

\begin{document}

\begin{frontmatter}

\title{Discovery Potential of Future Electron-Positron Colliders for a 95\,GeV Scalar}

\author{Pramod Sharma\fnref{inst1,inst2}}
\ead{pramodsharma.iiser@gmail.com}

\author{Anza-Tshilidzi Mulaudzi\fnref{inst1,inst3}}
\ead{anza-tshilidzi.mulaudzi@cern.ch}

\author{Karabo Mosala\fnref{inst1,inst3}}
\ead{karabo.mosala@cern.ch}

\author{Thuso Mathaha\fnref{inst1,inst3}}
\ead{thuso.mathaha@cern.ch}

\author{Mukesh Kumar\fnref{inst1}}
\ead{mukesh.kumar@cern.ch}

\author{Bruce Mellado\fnref{inst1,inst3}}
\ead{bmellado@mail.cern.ch}

\author{Andreas Crivellin\fnref{inst5}}
\ead{andreas.crivellin@psi.ch}

\author{Maxim Titov\fnref{inst6}}
\ead{maksym.titov@cern.ch}

\author{Manqi Ruan\fnref{inst7}}
\ead{manqi.ruan@ihep.ac.cn}

\author{Yaquan Fang\fnref{inst7,inst8}}
\ead{yaquan.fang@cern.ch}

\address[inst1]{School of Physics and Institute for Collider Particle Physics, University of the Witwatersrand, Johannesburg, Wits 2050, South Africa.}

\address[inst2]{Indian Institute of Science Education and Research, Knowledge City, Sector 81, S. A. S. Nagar, Manauli PO 140306, Punjab, India.}

\address[inst3]{iThemba LABS, National Research Foundation, PO Box 722, Somerset West 7129, South Africa}


\address[inst5]{Physik-Institut, Universit\"at Z\"urich, Winterthurerstrasse 190, CH--8057 Z\"urich, Switzerland}

\address[inst6]{
Commissariat à l’Énergie Atomique et Énergies Alternatives (CEA) Saclay
Direction de la Recherche Fondamentale (DRF)
Institute of Research into the Fundamental Laws of Universe (IRFU)
91191 Gif sur Yvette Cedex, France}

\address[inst7]{Institute of High Energy Physics, Chinese Academy of Sciences, Beijing, 100049, China}

\address[inst8]{University of Chinese Academy of Sciences, Beijing, 100049, China}

\begin{abstract}
The Large Electron Positron collider observed an indication for a new Higgs boson with a mass around $95$\,GeV-$100$\,GeV in the process $e^+e^-\to Z^*\to ZS$ with $S\to b\bar b$. The interest in this excess re-emerged with the di-photon signature at $\approx$\,95\,GeV at the Large Hadron Collider. In fact, a combined global significance of $3.4\sigma$ is obtained once $WW$ and $\tau\tau$ signals are included in addition. In this article, we perform a feasibility study for discovering such a new scalar $S$ at future electron-positron colliders using the recoil-mass method applied to $e^{+} e^{-} \to ZS$ with $Z \rightarrow \mu^{+} \mu^{-}$ and $S \to b  \bar{b}$. For this, we employ a Deep Neural Network to enhance the separation between the Standard Model background and the signal, reducing the required integrated luminosity necessary for discovery by a factor of two to three. As a result, an $SU(2)_L$ singlet Higgs with a mass of $\approx$\,95\,GeV can be observed with more than 5$\sigma$ significance at a 250\,GeV centre-of-mass energy collider with $5~ {\rm ab}^{-1}$ integrated luminosity if it has a mixing angle of at least $0.1$ with the Standard Model Higgs, which means that a discovery can be achieved within the whole 95\% confidence-level region preferred by Large Electron Positron excess. Furthermore, including more decay channels such as $S\to \tau\tau$ and $Z\to e^+e^-$ further enhances the discovery potential of future $e^+e^-$ accelerators, like CEPC, CLIC, FCC-ee and ILC.
\end{abstract}


\end{frontmatter}

\section{Introduction}
\label{sec:Intro}

The Large Electron-Positron (LEP) Collider at CERN reported in 2003 a mild excess with a local significance of 2.3$\sigma$ in the search for a new Higgs boson ($S$) in the range of 95\,GeV-100\,GeV using the process $e^+e^-\rightarrow ZS$ with $S\to b\bar b$~\citep{LEPWorkingGroupforHiggsbosonsearches:2003ing}, called Higgsstrahlung. Renewed interest in this excess emerged in 2018 when the CMS experiment released an analysis with an excess at $\approx 95$\,GeV in the di-photon invariant mass spectrum based on run-1 and partial run-2 data~\citep{CMS:2018cyk}. This result was later updated to include the full run-2 data, resulting in local (global) significance of 2.9$\sigma$ (1.3$\sigma$)~\citep{CMS:2024yhz}. Furthermore, ATLAS reported a smaller but consistent excess of 1.7$\sigma$ (locally) at the same mass~\citep{ATLAS:2024bjr} in di-photon final states and a 2.8$\sigma$ indication in the search for additional Higgs bosons in $\tau^{+} \tau^{-}$ was found by CMS~\citep{CMS:2022goy}.\footnote{Note that even though there is no dedicated ATLAS search in this channel, the side-band in the measurement of the Higgs boson cross-section in $\tau^{+} \tau^{-}$ final state~\citep{ATLAS:2022yrq} shows no excess, resulting in a reduction of the significance by approximately a factor $\sqrt 2$. Furthermore, also CMS finds no excess in $b$-associated production of a Higgs in the di-tau channel at around 95\,GeV.} In addition, an $\approx 2.5\sigma$ excess (at $\approx 95$\,GeV) has been found in the $W^{+} W^{-} \rightarrow \ell^{+} \ell^{-} \nu \nu, \ell = e, \mu$ channel~\citep{Coloretti:2023wng} by recasting the corresponding ATLAS and CMS SM Higgs analyses~\citep{CMS:2022uhn,ATLAS:2022ooq}. 

Therefore, we examine in this article the discovery prospects for a 95\,GeV Higgs boson at future $e^+e^-$ colliders. At these colliders, a scalar $S$ with a $m_S\approx 95$\,GeV can be produced via $e^{+} e^{-}\to Z^*\to Z S$ and the decay modes $Z \to \mu^{+} \mu^{-}$ and $S \to b  \bar{b}$ result in the most prominent signature. 
For our analysis, we consider the background process $e^{+} e^{-} \to Z b \bar{b}$ with $Z \to \mu^{+} \mu^{-}$ and use a deep neural network to enhance the discrimination of the signal (Higgs-strahlung process) from the background by employing the recoil mass method.
\begin{figure}[t]
\centering
\includegraphics[width=0.46\textwidth]{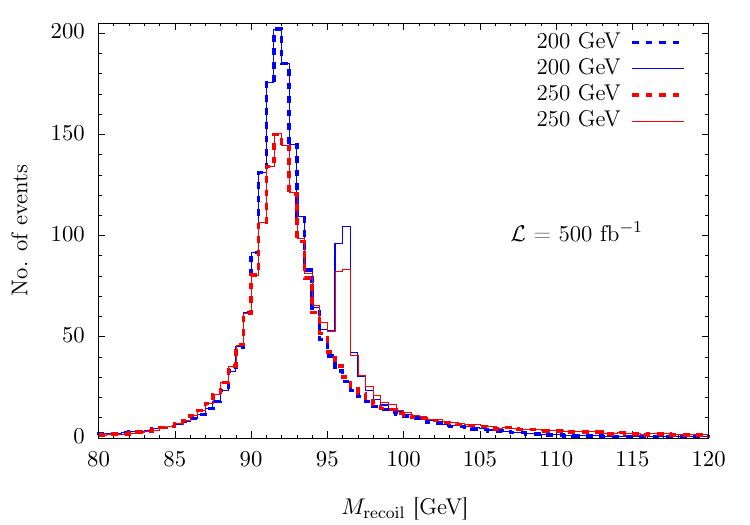}
\caption{Recoil-mass distribution of the simulated SM background (dashed) and the simulated signal plus background (solid) for $m_{S} = 95.5$\,GeV for  $\sqrt{s}=200$\,GeV (blue) and $\sqrt{s}=250$\,GeV (red) at $\mathcal{L}= 500$fb$^{-1}$.}
\label{fig:Rec}
\end{figure}

\begin{table*}[t]
    \centering
    \begin{tabular}{c|c|c|c}
    \hline
       Kinematic cuts  & $\mathcal{N}_S$  & $\mathcal{N}_B$ &  significance \\
       \hline
       \hline
    At least 2 $b$-tagged jets and 2 muons &  188 (219)  & 2926 (1876) & 2.4$\sigma$ (3.8$\sigma$)\\
    $E_{b,\mu} > $ 5\,GeV  & 188 (219)  & 2924 (1874) & 2.4$\sigma$ (3.8$\sigma$) \\
    $M_{b\bar{b}} < $ 100 GeV &   187 (217) & 2223 (1864) & 2.9$\sigma$ (3.8$\sigma$) \\
    $M_{\text{recoil}} < $ 120 GeV & 184 (216) & 1732 (1851) & 3.4$\sigma$ (3.8$\sigma$) \\ \hdashline
   93.5\,GeV $ < M_{\text{recoil}} <  $ 97.5\,GeV (before DNN) & 150 (193)  &  288 (274) & 8.4$\sigma$ (11.1$\sigma$)\\ 
   93.5\,GeV $ < M_{\text{recoil}} <  $ 97.5\,GeV (after DNN) &  54 (71) &  14 (14) &  14.5$\sigma$ (18.9$\sigma$) \\
    \hline
    \end{tabular}
    \caption{Cut-flow table showing the number of events for the signal ($\mathcal{N}_S$) and background ($\mathcal{N}_B$), along with the significance for the signal (as defined in \autoref{sig_sys}), after each kinematic cut for $\sqrt{s}$ = 250\,GeV (200\,GeV) at $\mathcal{L}$ = 500~$\text{fb}^{-1}$. The signal and background cross-sections are 1.1\,fb (1.4\,fb) and 16.8\,fb (12.79\,fb), respectively, for $\sqrt{s}$ = 250\,GeV (200\,GeV). Here, the signal cross-section for the 95.5\,GeV scalar is obtained by rescaling the one of a hypothetical 95.5\,GeV SM Higgs-boson by $\kappa^2_Z$ = 0.1.}
    \label{tab:cutflow}
\end{table*}

\section{Analysis and Results}\label{sec:results}

Higgs-strahlung, i.e.~$e^{+} e^{-} \to Z^* \to Z S$, is the dominant Higgs boson production mechanism at center-of-mass (c.o.m.) energies of $220\,{\rm GeV} \lesssim \sqrt{s} \lesssim 350$\,GeV while Higgs production through $WW$ and $ZZ$ fusion is subleading for SM-like Higgses~\citep{Chen:2016zpw}. It allows for a precise measurement of the $SZZ$ coupling in a model-independent approach, the `$\kappa$-framework' where the $SZZ$ coupling is parametrized by a factor $\kappa_{Z}$ such that $\kappa_{Z}=1$ for the SM Higgs. A measurement of $\kappa_{Z}$ can be performed at lepton colliders using the $Z$-boson recoil mass, specifically for events where $Z \to \mu^+\mu^-$. Since muons can be well identified, along with their momenta measured in the detector, Higgs-strahlung events can be identified using the recoil-mass method by tagging muon pairs~\citep{Chen:2016zpw}. The recoil mass is defined as: 
\begin{align}
\label{eq:3.2}
   M_{\rm recoil} = \sqrt{s + M_{\mu^{+} \mu^{-}}^2 - 2(E_{\mu^{+}} + {E_{\mu^{-}}})\sqrt{s}}\,,
\end{align}
such that reconstructing the Higgs boson mass is not required. Here, $M_{\mu^{+} \mu^{-}}$ is the invariant mass of the muon pair, and $E_{\mu^{+}}$ and $E_{\mu^{-}}$ are the energies of the positively and negatively charged muons, respectively. 

\subsection{Signal and background discrimination}
\label{subsec:DNN}

For the signal of a new Higgs boson $S$, we simulated the Higgs-strahlung process $e^+ e^- \to Z^* \to Z S$ with $Z \to \mu^+ \mu^-$ and $S \to b \bar{b}$, as well as the background process $e^{+} e^{-} \to Z b \bar{b}$ with $Z \to \mu^{+} \mu^{-}$, at a future electron-positron collider by generating one million events. These simulations were conducted for $m_S = 95.5$\,GeV scalar at center of mass (c.o.m.) energies of $ \sqrt{s} = 200$\,GeV and $\sqrt{s} = 250$\,GeV, using the Monte Carlo event generator {\tt MadGraph5}~\citep{Alwall:2011uj}. The BSM signal is assumed to be 10\% of the production cross-section of a (hypothetical) SM-like Higgs-boson with $m_S = 95.5$\,GeV in the following (if not defined otherwise). The generation-level cuts were applied to the transverse momentum $(p_{T\mu})$, pseudo-rapidity $(\eta_\mu)$ of the muons, and the angular distance between the muons ($\Delta R_{\mu^+ \mu^-}$), as $p_{T\mu} > 10$\,GeV, $|\eta_\mu| < 2.5$, and $\Delta R_{\mu^+ \mu^-} > 0.4$, respectively.

Following this, we performed the showering, fragmentation, and hadronization of the events with {\tt{Pythia8}}~\citep{Bierlich:2022pfr} and incorporated the detector simulation via {\tt{Delphes}}~\citep{deFavereau:2013fsa} based on the proposed CEPC detector  design~\citep{CEPCStudyGroup:2023quu}. Jets are clustered by {\tt{FastJet}} with the anti-$k_T$ algorithm using the distance parameter $R = 0.4$ and a dynamic scale both for the factorization and the renormalization.

To optimize the ratio of signal over background, we require exactly two muons and two $b$-tagged jets and apply minimal cuts on the energy of the $b$-tagged jets and muon of $E_{b,\mu} > 5$\,GeV.  Additionally, we use \autoref{eq:3.2} to compute the recoil mass for both signal and background shown in \autoref{fig:Rec}, and impose $M_{\mathrm{recoil}} < 120$\,GeV and $M_{b\bar{b}} < 100$\,GeV cuts on signal and background events for $\sqrt{s}=$ 250 (200) GeV. Out of the one-million unweighted events generated in {\tt MadGraph5}, 330,569 (307,988) signal events and 206,201 (289,502) background events survive the cuts for $\sqrt{s} = 250$\,GeV (200\,GeV). These optimization cuts enhance the signal-to-background ratio, defined over the whole $M_{\rm recoil}$ range, increasing the significance from 2.4$\sigma$ to 3.4$\sigma$ at $\mathcal{L}$ = 500 $\text{fb}^{-1}$ for $\sqrt{s} = 250$\,GeV for $\kappa_Z^2=0.1$ (before using the DNN) as summarized in \autoref{tab:cutflow}. As shown in \autoref{fig:Rec}, the $M_{\rm recoil}$ distribution exhibits a sharper peak around $m_S$ at $\sqrt{s} = 200$\,GeV compared to $\sqrt{s} = 250$\,GeV. This improvement can be attributed to the dependence of $M_{\rm recoil}$ on $M_{\mu^+ \mu^-}$ reconstruction, which performs better at lower c.o.m.~energies due to reduced boost effects.

\begin{figure*}[t]
\centering
\subfloat[\label{fig:significance}]{\includegraphics[width=0.45\textwidth,height=0.43\textwidth]{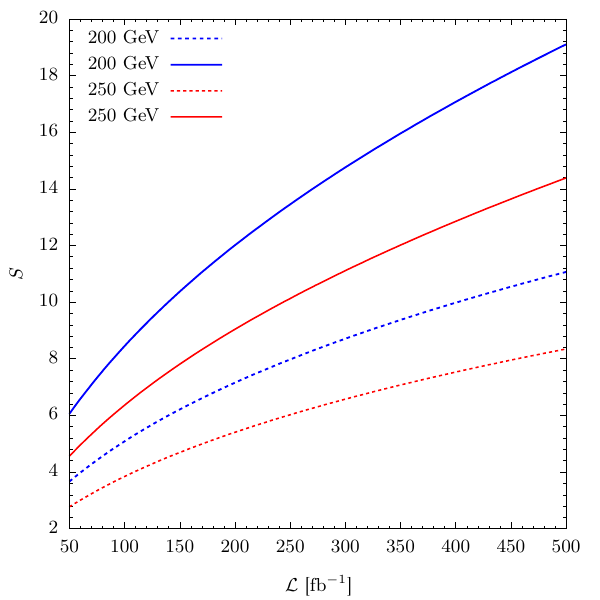}} \qquad
\subfloat[\label{fig:discovery}]{\includegraphics[width=0.45\textwidth,height=0.44\textwidth]{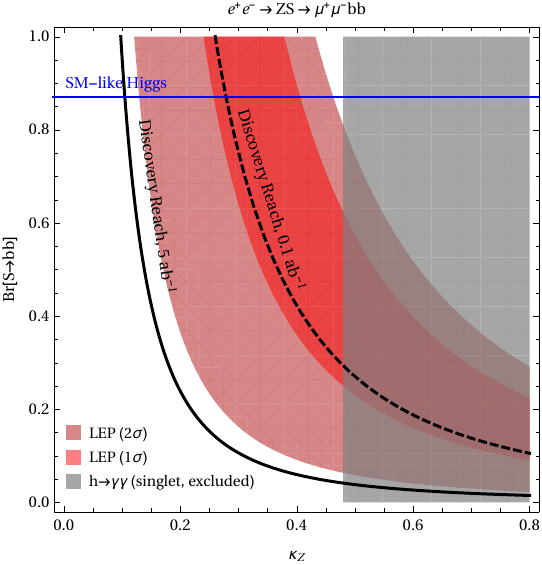}}
\caption{(a) The significance as a function of luminosity for $m_{S} = 95.5$\,GeV and $\sqrt{s}$ = 250\,GeV (200\,GeV) is shown in red (blue). We consider the number of events in the recoil mass window of 93.5\,GeV -- 97.5\,GeV before (after) the DNN has been applied as a dashed (solid) line. A cut of 0.965 (0.959) is applied on DNN response for $\sqrt{s}$ = 250\,GeV (200\,GeV). (b) Discovery region for a 95\,GeV scalar in the  $\kappa_Z$-Br($S\to b\bar b$) plane for an integrated luminosity of $\mathcal{L} = 5$ab$^{-1}$ at $\sqrt{s}=250$\,GeV. The red region is currently preferred by the LEP measurement of Higgs-strahlung. The blue line indicates the Br($S\to b\bar b$) for a SM-like Higgs, and the gray region is excluded by the di-photon signal strength of the SM Higgs assuming that $S$ is an $SU(2)_L$ singlet.}
\end{figure*}

{\it Machine Learning model:} For performing the machine learning (ML) analysis to improve on the discrimination between signal and background, we use a Deep Neural Network (DNN) which is structured as a Sequential Neural Network. It has a linear stack of layers, beginning with an input layer of 13 features (kinematic observables) reflecting the dimension of input data. These features include the azimuthal and polar angles of $b$-tagged jets and muons, the energies of $b$-tagged jets and the positively charged muon, the invariant mass of the muons, and the recoil mass of the muons. The architecture includes six hidden layers with 128, 64, 48, 32, 24, 16 and 8 neurons. Each layer uses the Rectified Linear Unit (ReLU) activation function~\citep{Agarap:2018uiz})
\begin{equation}
    f(x) \equiv{ \rm max} (0, x) = \begin{cases} 
      0 & \text{if } x < 0 \\
      x & \text{if } x \geq 0 
   \end{cases},
\end{equation}
where $x$ is the input variable.\footnote{Note that in our analysis, input variables have different scales and units. While we do not normalize input features before training, we employ batch normalization to normalize the mini-batches dynamically during training. This takes into account differences in scale.} It initializes weights for the layers using the HeUniform initializer~\citep{he2015delving}, which draws them from a uniform distribution within the range $[-\sqrt{6/{n_i}}, \sqrt{6/{n_i}}]$ based on the number of input neurons $n_i$ connected to the neuron for which weight is initialized. Each layer is followed by a batch normalization to stabilize training and accelerate convergence.\footnote{Batch normalization stabilizes and accelerates neural network training by standardizing layer inputs, reducing internal covariate shifts, and minimizing variation in input distributions, leading to faster convergence, higher learning rates, and less risk of vanishing or exploding gradients.} The model concludes with an output layer containing a single neuron and a {\tt{sigmoid}} activation function~\citep{narayan1997generalized}
\begin{equation}
    \sigma(x) = \frac{1}{1+e^{-x}},\quad {\rm with}\quad 0 < \sigma(x) <1.
\end{equation}
Here, $x$ is the input variable for the output layer which is the weighted sum of outputs from the previous layer. The weight in the output layer is initialized using the GlorotUniform initializer~\citep{pmlr-v9-glorot10a}. The model is compiled using the \texttt{Adam} optimizer~\citep{kinga2015method} with a learning rate of 0.0001. During compilation, the model uses binary cross-entropy~\citep{bishop2006pattern} with the loss function defined as
\begin{equation}
    -f(y_{\rm true},y_{\rm pred}) = y_{\rm true}\; \log(y_{\rm pred}) + (1 - y_{\rm true})\; \log(1 - y_{\rm pred}), 
\end{equation}
where $y_{\rm pred}$ is the probability that the DNN assigns to an event being signal or background, and $y_{\rm true}$ is the actual class label (1 for signal, 0 for background). 

With this ML model, we can improve the exploitation of the data to achieve better discrimination between signal and background events, i.e.~improve the event classification. For this purpose, the event samples remaining after the optimization cuts are split, with 70\% of the events allocated for training the DNN and the remaining 30\% reserved for testing its performance. Training is conducted with a batch size of 250 events from shuffled samples of signal and background at both $\sqrt{s}$ = 250\,GeV and $\sqrt{s}$ =200\,GeV. Although the network has a relatively small architecture, it is necessary to check whether it over-fits the training sample by looking at the DNN response, and the probability assigned to each event of being signal or background.

We plot the normalized number of events to which a specific value of the DNN model response is assigned as well as both signal and background events separately for the training and testing samples for $\sqrt{s}$ = 250\,GeV (200\,GeV). A significant difference is observed between the distributions of the DNN output obtained for the signal and background events. The DNN response demonstrates the classification accuracy of 89.1 (92.3)\% for $\sqrt{s}$ = 250\,GeV (200\,GeV) in training. The Area Under Curve (AUC) scores for $\sqrt{s}$ = 250\,GeV (200\,GeV), representing the DNN's ability to separate signal and background events across the dataset, are found to be 95.2 (97.4)\% and 95 (97.2)\% for training and testing samples, respectively. This consistency between training and testing AUC scores suggests that the model generalizes well, capturing signal and background patterns without over-fitting. 

\subsection{Signal Significance}\label{sec:sig}
The signal significance in this study is determined by the following formula
\begin{equation}
    S (\delta_{\rm sys}) = \frac{{\cal N}_\text{S}}{ \sqrt{{\cal N}_\text{B} + (\delta_{\rm sys} \cdot {\cal N}_\text{B})^{2}}}. \label{sig_sys}
\end{equation}
Here, $\mathcal{N}_\text{S}$ and $\mathcal{N}_\text{B}$ represent the number of signal and background events, respectively, at a given luminosity $\mathcal{L}$. The term $\delta_{\rm sys}$ accounts for systematic uncertainties in the measurement. The number of signal events is defined as $\mathcal{N}_\text{S} = \sigma_{\text{S}} \times \mathcal{L}$, where $\sigma_{\text{S}}$ represents the cross-section of the BSM signal. Similarly, the number of background events is given by ${\mathcal{N}_{\text{B}}} = \sigma_{\text{B}} \times \mathcal{L}$, where $\sigma_{\text{B}}$ denotes the cross-section of the SM background.

In \autoref{fig:significance}, we show $S (\delta_{\rm sys} = 2\%)$ both before and after DNN classification (with recoil mass requirements), for $\sqrt{s}$ = 250\,GeV (200\,GeV). The analysis is conducted within the range of $93.5\,{\rm GeV} \leq M_{\rm recoil} \leq 97.5$\,GeV. Comparatively, significance is enhanced after DNN classification. As shown in \autoref{tab:cutflow} for $\mathcal{L}$ = 500 $\text{fb}^{-1}$, a significance of approximately $8\sigma$ ($11\sigma$) can be achieved at a centre-of-mass energy of 200\,GeV (250\,GeV) before applying the DNN. After applying the DNN, the significance increased to 14$\sigma$ (19$\sigma$).

\subsection{Discovery potential}

We can now consider the discovery prospects for the scalar $m_S \approx 95$\,GeV at future $e^+e^-$ colliders. This can be done in a model-independent way by considering its branching ratio to $b\bar{b}$ as well as its coupling to $Z$-boson ($\kappa_Z$). In \autoref{fig:discovery} we present the regions in the $\kappa_Z -$Br($S \to b \bar{b}$) plane for $\sqrt{s} = 250$\,GeV, $\mathcal{L} = 0.1$\,ab$^{-1}$ and 5\,ab$^{-1}$ where a discovery is possible. Additionally, we show the preferred region from the analogous LEP measurement and the exclusion from $h\to\gamma\gamma$ in case $S$ is SM-like, i.e.~an $SU(2)_L$ singlet which obtains its couplings from mixing with the SM Higgs. Importantly, one can see that a discovery is possible within the whole 95\% confidence-level region preferred by the LEP excess. Note that the sensitivity can be further improved by including additional channels such as $S\to \tau^+\tau^-$~\cite{Dong:2025orv}, $S\to \gamma\gamma$, and $Z\to e^+e^-,\,\mu^+\mu^-$.

Note that for simplicity we consider here the case of a real $SU(2)_L$ singlet scalar. While this field alone is not sufficient to explain the excesses observed by the ATLAS and CMS experiments in the $\gamma\gamma$, $\tau^+\tau^-$, and $b\bar{b}$ channels, it can be part of an extended model that accounts for them. This is in particular the case for the 2HDM+S model~\cite{vonBuddenbrock:2016rmr, Muhlleitner:2016mzt}, where the 95\,GeV Higgs is mostly singlet-like~\cite{Azevedo:2023zkg}. In this setup, the $Z$-strahlung production cross section at future $e^+e^-$ colliders is still largely induced by the mixing with the SM-like Higgs doublet, such that our results remain valid to a good approximation. A similar study in the $S\to \tau^+\tau^-$ channel has been considered in Ref.~\cite{Dong:2025orv}, which also explores the viable parameter space in the 2HDM+S-flipped model.

\section{Conclusions and Outlook}
\label{sec:summary}

In this article, we explore the potential for the production of an additional scalar particle $S$ within the mass range of $m_{S}\approx$ 95~GeV in an $e^{+} e^{-}$ collider. Specifically, our investigation focuses on the Circular Electron-Positron Collider (CEPC) with center-of-mass energies of $\sqrt{s} = 200\,$GeV and $250$\,GeV and an integrated luminosity of $\mathcal{L} = 500$~fb$^{-1}$. The choice of utilizing ML techniques, particularly DNNs, is motivated by their proven accuracy in event classification.

We first demonstrated the advantage of employing DNNs by examining the reconstruction of the scalar mass, $m_S$, using the recoil mass method, as depicted in \autoref{fig:Rec}. The effectiveness of DNNs in enhancing signal significance is evident from~\autoref{fig:significance}, where a substantial improvement in the signal significance is observed. To be more specific, using the DNN a significance of $\approx8\sigma$ ($\approx11\sigma$) for a new scalar with a mass of 95.5\,GeV and $\kappa_Z^2=0.1$ can be achieved at a c.o.m.~energy of \,GeV (250\,GeV) for ${\cal {L}} = 500$~fb$^{-1}$, which corresponds to an increase of 75\% (73\%). This highlights the critical role of ML techniques in enhancing the sensitivity of collider measurements. 

Electron-positron Higgs factories offer a unique capability to tag the $e^{+} e^{-} \to Z S$ signal using the recoil mass method, independent of the Higgs boson decay models. Taking the CEPC as an example, which is expected to deliver 5~$\rm ab^{-1}$ at a c.o.m. energy of 240\,GeV -- 250\,GeV over 10~years of data taking~\cite{Chen:2016zpw,CEPCStudyGroup:2023quu}, it could identify the recoil mass signal of $S$ within approximately 100 days and lead to a 5$\sigma$ discovery. However, as analyzed in this article, applying an ML technique reduces the time required to achieve this discovery potential to about 34~days. Note that in this analysis, we considered a tagging efficiency of 80\% for $b$-tagged jets, based on the proposed CEPC detector design. However, this efficiency could be improved to as high as 92\%, as studied in Ref.~\citep{PhysRevLett.132.221802}, which would further enhance the significance and discovery potential of the 95\,GeV signal. Additionally, this analysis only considers the $Z \to \mu^+ \mu^-$ decay channel; including other decay modes such as $Z \to e^+ e^-$ and $Z \to \tau^+ \tau^-$ would further improve the results.

\bibliography{cas-refs}

\end{document}